\documentclass[draftcls, onecolumn]{IEEEtran}

\usepackage{graphicx}
\usepackage[cmex10]{amsmath}
\usepackage{amsfonts}
\usepackage{cite}
\usepackage{url}
\usepackage{soul}

\providecommand{\Tensor}[1]{\bar{\bar {#1}}} % tensor, matrix

\newcommand{\BE}{\begin{equation}}   % begin of a equation
\newcommand{\EE}{\end{equation}}     % end of a equation
\newcommand{\BF}{\begin{figure}}     % begin of a figure
	\newcommand{\EF}{\end{figure}}  	 % -//-

\author{
	\IEEEauthorblockN{Lukas~Jelinek\IEEEauthorrefmark{1}, Ondrej~Kratky\IEEEauthorrefmark{2}, Miloslav~Capek\IEEEauthorrefmark{1} \\}
	\IEEEauthorblockA{\IEEEauthorrefmark{1}Department of Electromagnetic Field, Faculty of Electrical Engineering, Czech Technical University in Prague, Technicka 2, 16627, Prague, Czech Republic, Email: lukas.jelinek@fel.cvut.cz \\}
	\IEEEauthorblockA{\IEEEauthorrefmark{2}Siemens Czech Republic Ltd, Siemensova 1, 15500, Prague}
}

\title{An Evaluation of Polarizability Tensors of Arbitrarily Shaped Highly Conducting Bodies}

\begin{document}
	
	\maketitle
	
	\begin{abstract}
	A full-wave numerical scheme of  polarizability (polarisability)  tensors evaluation is presented. The method accepts highly conducting bodies of arbitrary shape and explicitly accounts for the radiation as well as ohmic losses. The method is verified on canonical bodies with known polarizability tensors, such as a sphere and a cube, as well as on realistic scatterers. The theoretical developments are followed by a freely available code whose sole user input is the triangular mesh covering the surface of the body under consideration. 
	\end{abstract}
	
	%\begin{IEEEkeywords}
	%XXXXX
	%\end{IEEEkeywords}
%=============================================================================================================================================================================

%%%%%%%%%%%%%%%%%%%%%%%%%%%%%%%%%%%%%%%%%%%%%%%%%%%%%%%%%%%%%%%%%%%%%%%%%%%%%%%%%%%%%%%%%%%%%%%%%%%%%%%%%%%%%%%%%%%%%%%%%%%%%%%%%%%%%%%%%%%%%%%%%%%%%%%%%%%%%%%%%%%%%%%%%%%%%%
\section{Introduction}
\label{Intro}
Polarizability tensors \cite{Bethe_Theory_of_Diffraction_by_Small_Holes,Bouwkamp_Diffraction_Theory,Collin_FieldTheoryOfGuidedWaves,Tretyakov2003} are an indispensable tool for designing artificial materials \cite{Simovski2011,Simovski2011a} and frequency selective surfaces \cite{Wu_1995,Munk_FrequencySelectiveSurfaces}. Thanks to the relation of the polarizability tensors to the radar cross-section \cite{SihvolaSarkarKolundzija2004} or scattering cross-section \cite{Bohren_scattering_by_Small_Particles,Jackson_ClassicalElectrodynamics}, they also present a vital tool for designing radiofrequency identification (RFID) tags \cite{Finkenzeller_RFID}. Last, but not least, the polarizability tensors also fully characterize the radiation properties of electrically small antennas \cite{GustafssonSohlKristensson_PhysicalLimitationsOfAntennasOfArbitraryShape_RoyalSoc,YaghjianStuart_LowerBoundOnTheQofElectricallySmallDipoleAntennas}. The precise evaluation of the polarizability tensors is, thus, of major interest for many branches of applied electromagnetism.

In canonical cases, there exist analytical models for polarizability \cite{Collin_FieldTheoryOfGuidedWaves,Tretyakov2003}. However, during the development of metamaterials \cite{MarquesMartinSorolla2007,SolymarShamonina2009}, chipless RFID tags \cite{PreradovicKarmakar2010} and modern reflection~/~transmission arrays \cite{Pfeiffer_2013_PRL}, the geometry of their basic constituents become complex and the polarizability tensors of realistic scatterers can only be extracted through  numerical methods. Early attempts \cite{Mei_1963,Meulenaere_1977,Ercument_1983} and some of their extensions \cite{Avelin_2001,Sihvola_2004,Helsing_2013} were purely static in nature while ignoring important \cite{Pendry2004} magnetoelectric coupling and radiation losses. Modern approaches rely on commercial full-wave electromagnetic solvers which evaluate the induced currents on a scatterer \cite{IshimaruLeeKugaEtAl2003,2016_Yazdi_PIERM} or the scattered far-fields \cite{2014_Asadchy_PNFA}. The usage of powerful commercial packages makes it possible to work with complex scatterers including non-reciprocal materials \cite{2014_Asadchy_PNFA}. Unfortunately, such generality is encumbered with higher computational demands, making these schemes time-consuming which is especially problematic in conjunction with structural optimization.

Apart from the direct numerical evaluation of polarizability tensors, attempts have also been to measure them. Pioneering work in this direction has been done by Cohn \cite{Cohn1952}, in which the scatterer is placed in an electrolyte. For recent methods, using a vacuum environment, we consider \cite{ReinertJacob2001,JelinekBaenaMarquesEtAl2006,Jelinek_2014_AWPL} which describe how to obtain a particle's polarizability by measuring the scattering parameters of a waveguide segment loaded by the analysed body, or \cite{ScherKuester2009,2012_Albooyeh_PRB,2013_Karamanos_IETMAP,2016_Liu_TAP} which use a measurement of the scattering parameters of a two-dimensional (2D) array of analysed bodies. These methods can deal with scatterers of quite general shape and constitution but generally suffer from the necessity of removing the effect of the artificial periodic environment or the waveguide walls. This is done either by a suitable calibration process \cite{Jelinek_2014_AWPL,2016_Liu_TAP}, or by directly evaluating the interaction constants of the array
\cite{ScherKuester2009,2012_Albooyeh_PRB,2013_Karamanos_IETMAP}. The basic deficiency of the calibration procedure is the inaccesibility of precisely defined bianistropic standards. The direct evaluation of interaction constants is not without problems either. The fundamental issue is that a point dipole approximation is used when making the problem mathematically tractable \cite{BelovSimovski2005,Tretyakov2003,ScherKuester2009} and this introduces systematic errors. These errors decay with the sparsity of the lattice, but sparse lattices introduce numerical errors due to very low reflectivity and possible higher order reflected and transmitted modes. There is, thus, a necessary trade-off between the aforementioned errors, which is generally shape-dependent. 
 
In this paper we propose and verify a general method to extract of all four polarizability tensors of arbitrarily shaped bodies with finite conductivity. The presented scheme uses full-wave numerical evaluation, automatically accounting for ohmic and radiation losses. The paper also discusses  numerically efficient  implementation of this method in the Rao-Wilton-Glisson basis \cite{RaoWiltonGlisson_ElectromagneticScatteringBySurfacesOfArbitraryShape} which results in a freely available code \cite{polarizability_extraction_at_file_exchange}.

The paper is organized as follows. Section~\ref{method} introduces the polarizability extraction scheme. The method is verified in Section~\ref{results} and its most salient features are discussed in Section~\ref{disc}. The paper concludes in Section~\ref{concl}. Various derivations necessary for the implementation of the proposed method are presented in Section~\ref{app}.

%%%%%%%%%%%%%%%%%%%%%%%%%%%%%%%%%%%%%%%%%%%%%%%%%%%%%%%%%%%%%%%%%%%%%%%%%%%%%%%%%%%%%%%%%%%%%%%%%%%%%%%%%%%%%%%%%%%%%%%%%%%%%%%%%%%%%%%%%%%%%%%%%%%%%%%%%%%%%%%%%%%%%%%%%%%%%%
\section{Description of the Method}
\label{method}

\subsection{Definition of Polarizability Tensors}
\label{tensors}
 Let us assume an electrically small scatterer ($ka \ll 1$) fully enclosed in a sphere of radius $a$, centered in the coordinate system, with $k$ being the freespace wavenumber  \cite{Harrington_TimeHarmonicElmagField}. Under the assumtion of a time-harmonic steady state \cite{Harrington_TimeHarmonicElmagField}, i.e., \mbox{$\mathcal{F} \left( t \right)=\mathrm{Re} \left\{F \left( \omega \right) \mathrm{exp} \left(\mathrm{j} \omega t \right)\right\} $}, with angular frequency $\omega$, the illumination of the scatterer by an incident electromagnetic wave with electric field ${\boldsymbol{E}}^{\left[ {3 \times 1} \right]}\left( {\boldsymbol{r}} \right)$ and magnetic field ${\boldsymbol{B}}^{\left[ {3 \times 1} \right]}\left( {\boldsymbol{r}} \right)$ gives rise to electric and magnetic dipole moments ${{\boldsymbol{p}}^{\left[ {3 \times 1} \right]}},{{\boldsymbol{m}}^{\left[ {3 \times 1} \right]}}$ \cite{Jackson_ClassicalElectrodynamics} 
\BE
\label{2Aeq01}
\left[ {\begin{array}{*{20}{c}}
	{\boldsymbol{p}}\\
	{\boldsymbol{m}}
	\end{array}} \right] = \left[ {\begin{array}{*{20}{c}}
	{\boldsymbol{\Tensor{\alpha}}_{{\mathrm{ee}}}}&{\boldsymbol{\Tensor{\alpha}}_{{\mathrm{em}}}}\\
	{\boldsymbol{\Tensor{\alpha}}_{{\mathrm{me}}}}&{\boldsymbol{\Tensor{\alpha}}_{{\mathrm{mm}}}}
	\end{array}} \right]\left[ {\begin{array}{*{20}{c}}
	{{\boldsymbol{E}}\left( 0 \right)}\\
	{{\boldsymbol{B}}\left( 0 \right)}
	\end{array}} \right],
\EE
where $\boldsymbol{\Tensor{\alpha}}_{{\mathrm{ee}}}^{\left[ {3 \times 3} \right]},\;\boldsymbol{\Tensor{\alpha}}_{{\mathrm{em}}}^{\left[ {3 \times 3} \right]},\;\boldsymbol{\Tensor{\alpha}}_{{\mathrm{me}}}^{\left[ {3 \times 3} \right]},\;\boldsymbol{\Tensor{\alpha}}_{{\mathrm{mm}}}^{\left[ {3 \times 3} \right]}$ are the dipolar polarizability tensors \cite{Collin_FieldTheoryOfGuidedWaves,Tretyakov2003}. The polarizability tensors $\boldsymbol{\Tensor{\alpha}}$ are known \cite{Bohren_scattering_by_Small_Particles} to fully characterize the scattering properties of electrically small scatterers. Assuming a scatterer made of highly conductive material, the electric and magnetic dipole moments can be evaluated from the knowledge of the induced surface current density  ${{\boldsymbol{K}}^{\left[ {3 \times 1} \right]}\left( {\boldsymbol{r}} \right)}$ as 
\begin{subequations}
	\begin{align}
	\label{2Aeq02a}
	{\boldsymbol{p}} &= \frac{1}{{{\mathrm{j}}\omega }}\int\limits_S {{\boldsymbol{K}}\left( {\boldsymbol{r}} \right){\mathrm{d}}S},	\\
	\label{2Aeq02b}
	{\boldsymbol{m}} &= \frac{1}{2}\int\limits_S {{\boldsymbol{r}} \times {\boldsymbol{K}}\left( {\boldsymbol{r}} \right){\mathrm{d}}S},
	\end{align}
\end{subequations}
where $\boldsymbol{r}^{\left[ {3 \times 1} \right]}$ is a radius vector from a suitably chosen origin, commonly taken to coincide with the geometrical centre of a scatterer, see \ref{PMmatrices} for a commentary on a possible coordinate dependence.

\subsection{Evaluation of Polarizability Tensors}
\label{evaluation}
Imagine that we set up six different excitation scenarios
\BE
\label{2Beq01}
\left[ {\begin{array}{*{20}{c}}
	{\begin{array}{*{20}{c}}
		{{{\boldsymbol{E}}_1}\left( 0 \right)}\\
		{{{\boldsymbol{B}}_1}\left( 0 \right)}
		\end{array}}& \cdots &{\begin{array}{*{20}{c}}
		{{{\boldsymbol{E}}_6}\left( 0 \right)}\\
		{{{\boldsymbol{B}}_6}\left( 0 \right)}
		\end{array}}
	\end{array}} \right]
\EE
producing six polarizations  $\left[{{\boldsymbol{p}}_1} \cdots {{\boldsymbol{p}}_6}\right]$ and $\left[{{\boldsymbol{m}}_1} \cdots {{\boldsymbol{m}}_6}\right]$  of the scatterer. Assume further that the excitations are chosen to make columns of (\ref{2Beq01}) linearly independent. In such a case the polarizability tensors can be evaluated as
\BE
\label{2Beq02}
\left[ {\begin{array}{*{20}{c}}
	{\boldsymbol{\Tensor{\alpha}}_{{\mathrm{ee}}}^{}}&{\boldsymbol{\Tensor{\alpha}}_{{\mathrm{em}}}^{}}\\
	{\boldsymbol{\Tensor{\alpha}}_{{\mathrm{me}}}^{}}&{\boldsymbol{\Tensor{\alpha}}_{{\mathrm{mm}}}^{}}
	\end{array}} \right] = \left[ {\begin{array}{*{20}{c}}
	{\begin{array}{*{20}{c}}
		{{{\boldsymbol{p}}_1}}\\
		{{{\boldsymbol{m}}_1}}
		\end{array}}& \cdots &{\begin{array}{*{20}{c}}
		{{{\boldsymbol{p}}_6}}\\
		{{{\boldsymbol{m}}_6}}
		\end{array}}
	\end{array}} \right]{\left[ {\begin{array}{*{20}{c}}
		{\begin{array}{*{20}{c}}
			{{{\boldsymbol{E}}_1}\left( 0 \right)}\\
			{{{\boldsymbol{B}}_1}\left( 0 \right)}
			\end{array}}& \cdots &{\begin{array}{*{20}{c}}
			{{{\boldsymbol{E}}_6}\left( 0 \right)}\\
			{{{\boldsymbol{B}}_6}\left( 0 \right)}
			\end{array}}
		\end{array}} \right]^{ - 1}}.
\EE

Throughout this paper, the surface current density ${{\boldsymbol{K}}\left( {\boldsymbol{r}} \right)}$, needed for evaluation of (\ref{2Aeq02a}) and (\ref{2Aeq02b}), is obtained from the Electric Field Integral Equation (EFIE) \cite{Harrington_FieldComputationByMoM} (see Section~\ref{app} for details) discretized in a given basis 
\BE
\label{2Beq03}
{\boldsymbol{K}}\left( {\boldsymbol{r}} \right) \approx \sum\limits_n {{I_n}{{\boldsymbol{f}}_n}\left( {\boldsymbol{r}} \right)},
\EE
where ${{\mathbf{I}}^{\left[ {N \times 1} \right]}}$ is the vector of expansion coefficients and ${{\boldsymbol{f}}_n}\left( {\boldsymbol{r}} \right)$ are suitable real dimensionless basis functions. The expansion (\ref{2Beq03}) transforms the EFIE into
\BE
\label{2Beq04}
\left(\Tensor{\mathbf{\Sigma}}-\Tensor{\mathbf{Z}}\right) \mathbf{I} = \left[ {\begin{array}{*{20}{c}}
	{\left\langle {{{\boldsymbol{f}}_1},{\boldsymbol{E}}} \right\rangle }\\
	\vdots \\
	{\left\langle {{{\boldsymbol{f}}_N},{\boldsymbol{E}}} \right\rangle }
	\end{array}} \right],
\EE
with ${{\Tensor{\mathbf{Z}}}^{\left[ {N \times N} \right]}}$ as the well-known impedance matrix \cite{Harrington_FieldComputationByMoM}, with ${{\Tensor{\mathbf{\Sigma}}}^{\left[ {N \times N} \right]}}$ as the matrix representing the reaction of a lossy conductor and with
\BE
\label{2Beq05}
	\left\langle {{\boldsymbol{f}},{\boldsymbol{g}}} \right\rangle  = \int\limits_S {{\boldsymbol{f}}^*\left( {\boldsymbol{r}} \right) \cdot {{\boldsymbol{g}}}\left( {\boldsymbol{r}} \right){\mathrm{d}}S}
\EE
as a suitably defined scalar product. The construction of matrices $\Tensor{\mathbf{Z}}$ and $\Tensor{\mathbf{\Sigma}}$ is detailed in Section~\ref{app} and Section~\ref{Sigmamat}. Furthermore, substituting (\ref{2Beq03}) into (\ref{2Aeq02a}) and (\ref{2Aeq02b}), and utilizing (\ref{2Beq04}) allow us to write
\BE
\label{2Beq06}
\left[ {\begin{array}{*{20}{c}}
	{\boldsymbol{p}}\\
	{\boldsymbol{m}}
	\end{array}} \right] = \left[ {\begin{array}{*{20}{c}}
	{\Tensor{\mathbf{P}}}\\
	{\Tensor{\mathbf{M}}}
	\end{array}} \right]{\bf{I}} = \left[ {\begin{array}{*{20}{c}}
	{\Tensor{\mathbf{P}}}\\
	{\Tensor{\mathbf{M}}}
	\end{array}} \right]{\left( {\Tensor{\mathbf{\Sigma}}  - {\Tensor{\mathbf{Z}}}} \right)^{ - 1}}\left[ {\begin{array}{*{20}{c}}
	{\left\langle {{{\boldsymbol{f}}_1},{\boldsymbol{E}}} \right\rangle }\\
	\vdots \\
	{\left\langle {{{\boldsymbol{f}}_N},{\boldsymbol{E}}} \right\rangle }
	\end{array}} \right],
\EE
where matrices ${{\Tensor{\mathbf{P}}}^{\left[ {3 \times N} \right]}},{{\Tensor{\mathbf{M}}}^{\left[ {3 \times N} \right]}}$ are representations of (\ref{2Aeq02a}) into (\ref{2Aeq02b}) in the basis (\ref{2Beq03}). The construction of matrices $\Tensor{\mathbf{P}}$ and $\Tensor{\mathbf{M}}$ is detailed in Section~\ref{PMmatrices}.

Putting everything together, we have
\BE
\label{2Beq07}
\left[ {\begin{array}{*{20}{c}}
	{\boldsymbol{\Tensor{\alpha}}_{{\mathrm{ee}}}^{}}&{\boldsymbol{\Tensor{\alpha}}_{{\mathrm{em}}}^{}}\\
	{\boldsymbol{\Tensor{\alpha}}_{{\mathrm{me}}}^{}}&{\boldsymbol{\Tensor{\alpha}}_{{\mathrm{mm}}}^{}}
	\end{array}} \right] = \left[ {\begin{array}{*{20}{c}}
	{\Tensor{\mathbf{P}}}\\
	{\Tensor{\mathbf{M}}}
	\end{array}} \right]{\left( {\Tensor{\mathbf{\Sigma}}  - {\Tensor{\mathbf{Z}}}} \right)^{ - 1}}\left[ {\begin{array}{*{20}{c}}
	{\left\langle {{{\boldsymbol{f}}_1},{{\boldsymbol{E}}_1}} \right\rangle }& \cdots &{\left\langle {{{\boldsymbol{f}}_1},{{\boldsymbol{E}}_6}} \right\rangle }\\
	\vdots & \ddots & \vdots \\
	{\left\langle {{{\boldsymbol{f}}_N},{{\boldsymbol{E}}_1}} \right\rangle }& \cdots &{\left\langle {{{\boldsymbol{f}}_N},{{\boldsymbol{E}}_6}} \right\rangle }
	\end{array}} \right]{\left[ {\begin{array}{*{20}{c}}
		{\begin{array}{*{20}{c}}
			{{{\boldsymbol{E}}_1}\left( 0 \right)}\\
			{{{\boldsymbol{B}}_1}\left( 0 \right)}
			\end{array}}& \cdots &{\begin{array}{*{20}{c}}
			{{{\boldsymbol{E}}_6}\left( 0 \right)}\\
			{{{\boldsymbol{B}}_6}\left( 0 \right)}
			\end{array}}
		\end{array}} \right]^{ - 1}}.
\EE
The original complex scattering problem is now transformed into a trivial multiplication of matrices \cite{Harrington_MatrixMethodsForFieldProblems}. The only pending issue is to find six suitable excitations generated by ${{\boldsymbol{E}}_1 \left(\boldsymbol{r}\right)},\,...\,,\,{{\boldsymbol{E}}_6 \left(\boldsymbol{r}\right)}$. Note that magnetic fields ${{\boldsymbol{B}}_1 \left(\boldsymbol{r}\right)},\,...\,,\,{{\boldsymbol{B}}_6 \left(\boldsymbol{r}\right)}$ cannot be chosen freely as they are connected to the electric fields via freespace Maxwell's equations.

The raw form of (\ref{2Beq07}) is ill-suited for numerical implementation since the doublets $\Tensor{\mathbf{P}}$, $\Tensor{\mathbf{M}}$ and ${{\boldsymbol{E}}}\left( 0 \right)$, ${{\boldsymbol{B}}}\left( 0 \right)$ have different units and, consequently, considerably different magnitudes. For the sake of numerical stability it is then advantageous to use the following normalization
\BE
\label{3eq01}
\left[ {\begin{array}{*{20}{c}}
		{\boldsymbol{\Tensor{\alpha}}_{{\mathrm{ee}}}}&{\boldsymbol{\Tensor{\alpha}}_{{\mathrm{em}}}}\\
		{\boldsymbol{\Tensor{\alpha}}_{{\mathrm{me}}}}&{\boldsymbol{\Tensor{\alpha}}_{{\mathrm{mm}}}}
	\end{array}} \right] \to \left[ {\begin{array}{*{20}{c}}
	{\displaystyle\frac{{\boldsymbol{\Tensor{\alpha}}_{{\mathrm{ee}}}}}{{{\varepsilon _0}V}}}&\displaystyle{\frac{{{Z_0}\boldsymbol{\Tensor{\alpha}}_{{\mathrm{em}}}}}{V}}\\
	\displaystyle{\frac{{{Z_0}\boldsymbol{\Tensor{\alpha}}_{{\mathrm{me}}}}}{V}}&\displaystyle{\frac{{{\mu _0}\boldsymbol{\Tensor{\alpha}}_{{\mathrm{mm}}}}}{V}}
\end{array}} \right]
\EE

\BE
\label{3eq02}
\left[ {\begin{array}{*{20}{c}}
		{\Tensor{\mathbf{P}}}\\
		{\Tensor{\mathbf{M}}}
	\end{array}} \right] \to \left[ {\begin{array}{*{20}{c}}
	{\displaystyle\frac{{c_0\Tensor{\mathbf{P}}}}{{V}}}\\
	{\displaystyle\frac{{{\Tensor{\mathbf{M}}}}}{V}}
\end{array}} \right]
\EE

\BE
\label{3eq02a}
{\Tensor{\mathbf{\Sigma}}  - {\Tensor{\mathbf{Z}}}} \to \frac{1}{Z_0}\left({\Tensor{\mathbf{\Sigma}}  - {\Tensor{\mathbf{Z}}}}\right)
\EE

\BE
\label{3eq03}
\left[ {\begin{array}{*{20}{c}}
		{\begin{array}{*{20}{c}}
				{{{\boldsymbol{E}}_1}\left( 0 \right)}\\
				{{{\boldsymbol{B}}_1}\left( 0 \right)}
			\end{array}}& \cdots &{\begin{array}{*{20}{c}}
			{{{\boldsymbol{E}}_6}\left( 0 \right)}\\
			{{{\boldsymbol{B}}_6}\left( 0 \right)}
		\end{array}}
	\end{array}} \right] \to \left[ {\begin{array}{*{20}{c}}
	{\begin{array}{*{20}{c}}
			{{{\boldsymbol{E}}_1}\left( 0 \right)}\\
			{{c_0}{{\boldsymbol{B}}_1}\left( 0 \right)}
		\end{array}}& \cdots &{\begin{array}{*{20}{c}}
		{{{\boldsymbol{E}}_6}\left( 0 \right)}\\
		{{c_0}{{\boldsymbol{B}}_6}\left( 0 \right)}
	\end{array}}
\end{array}} \right]
\EE
with $V = 4\pi {a^3}/3$ being the volume of the smallest sphere circumscribing the scatterer, with $Z_0$ as the free-space impedance, $\varepsilon_0$ as the vacuum permittivity, $\mu_0$ as the vacuum permeability and $c_0$ as the speed of light. In that case, the polarizability matrix (\ref{3eq01}) and dipole moment matrix (\ref{3eq02}) become dimensionless, the matrix (\ref{3eq02a}) attains the dimension of $\mathrm{m}^2$ and, finally, the normalized excitation matrix (\ref{3eq03}) is in volts per meter.

\subsection{Excitation}
\label{excit}
An immediate candidate for the excitation is to have a normalized excitation matrix (\ref{3eq03}) equal to a unity matrix. The construction of this excitation is illustrated on the 3rd and 6th column of (\ref{3eq03}). To construct the 3rd column, choose cylindrical coordinates $\left(\rho,\varphi,z\right)$ and
\BE
\begin{aligned}
\label{2Ceq01}
{\boldsymbol{E}}\left( {\boldsymbol{r}} \right) &= {{\boldsymbol{z}}_0}{\mathrm{J}_0}\left( {k\rho } \right),\\
{c_0}{\boldsymbol{B}}\left( {\boldsymbol{r}} \right) &= \mathrm{j}{{\boldsymbol{\varphi }}_0}{\mathrm{J}_1}\left( {k\rho } \right),
\end{aligned}
\EE
where $\mathrm{J}_i$ represents Bessel's function of the first kind \cite{Arfken_MathForPhysicists}. Analogously, to construct the 6th column, choose
\BE
\begin{aligned}
\label{2Ceq02}
{\boldsymbol{E}}\left( {\boldsymbol{r}} \right) &=  -\mathrm{j} {{\boldsymbol{\varphi }}_0}{\mathrm{J}_1}\left( {k\rho } \right),\\
{c_0}{\boldsymbol{B}}\left( {\boldsymbol{r}} \right) &= {{\boldsymbol{z}}_0}{\mathrm{J}_0}\left( {k\rho } \right).
\end{aligned}
\EE
It can be checked that (\ref{2Ceq01}) and (\ref{2Ceq02}) satisfy the source-free Maxwell's equations in a vacuum and that for small values of $k \rho$ field (\ref{2Ceq01}) tends to

\BE
\begin{aligned}
\mathop {\lim }\limits_{k\rho  \to 0} {\boldsymbol{E}}\left( {\boldsymbol{r}} \right) &= {{\boldsymbol{z}}_0},\\
\mathop {\lim }\limits_{k\rho  \to 0} {c_0\boldsymbol{B}}\left( {\boldsymbol{r}} \right) &= 0,
\end{aligned}
\EE
while field (\ref{2Ceq02}) tends to
\BE
\begin{aligned}
\mathop {\lim }\limits_{k\rho  \to 0} {\boldsymbol{E}}\left( {\boldsymbol{r}} \right) &= 0,\\
\mathop {\lim }\limits_{k\rho  \to 0} {c_0\boldsymbol{B}}\left( {\boldsymbol{r}} \right) &= {{\boldsymbol{z}}_0},
\end{aligned}
\EE
exactly as desired. Other columns of (\ref{3eq03}) can easily be obtained by rotating (\ref{2Ceq01}) or (\ref{2Ceq02}).

The aforementioned excitation is simple to implement and Section~\ref{results} shows that it can be used in practice. Its utilization within the EFIE paradigm can, however, lead to odd behaviour at small electrical sizes for the EFIE solely uses electric field as an excitation, see (\ref{2Beq07}) and Section~\ref{app}. This poses no problem for electric type  excitation (\ref{2Ceq01}) (first three columns of (\ref{3eq03})), but can be problematic for magnetic type excitation (\ref{2Ceq02}) (last three columns of (\ref{3eq03})), which for vanishingly small electrical sizes generates no excitation at all.

Although the excitation described above is the first choice for its simplicity, the method described in Section~\ref{method} is not restricted to it and can be used with any other form of incident field, provided that: (a) the columns of the matrix (\ref{2Beq01}) are linearly independent; (b) the incident field satisfies freespace Maxwell's equations; and (c) its spatial variation can be neglected in a volume occupied by the scatterer under test. These conditions forbid the use of a homogeneous field, but allow a commonly employed excitation by linearly independent planewaves \cite{IshimaruLeeKugaEtAl2003}. In fact, due to identity \cite{Arfken_MathForPhysicists}
\BE
\label{2Ceq03}
\frac{1}{{2\pi }}\int\limits_0^{2\pi } {{{\mathrm{e}}^{{\mathrm{j}}\left( {x\cos \alpha  + y\sin \alpha } \right)}}{\mathrm{d}}\alpha } = {{\mathrm{J}}_0}\left( {k{\sqrt {{x^2} + {y^2}} } } \right)  = {{\mathrm{J}}_0}\left( {k\rho } \right),
\EE
the excitation (\ref{2Ceq01}) can be seen as an addition of uniformly angularly distributed planewaves which present a slight advantage over standalone planewaves by exhibiting rotationally symmetric amplitude and phase variation.

%=============================================================================================
%=============================================================================================
\section{Results}
\label{results}
To obtain numerical results, the method described in Section~\ref{method} has been implemented in Matlab \cite{matlab} using the RWG basis \cite{RaoWiltonGlisson_ElectromagneticScatteringBySurfacesOfArbitraryShape} in which the surface of the scatterer is decomposed into triangular patches and expansion coefficients $\mathbf{I}$ become the values of the RWG edge surface current densities \cite{RaoWiltonGlisson_ElectromagneticScatteringBySurfacesOfArbitraryShape}. The code used for the evaluation can be found at \cite{polarizability_extraction_at_file_exchange}. All results presented in this section are normalized according to (\ref{3eq01}).

The verification of the proposed extraction method starts with the polarizability of canonical bodies, namely a perfectly electrically conducting (PEC) sphere and cube, whose static polarizabilities are known either analytically~\cite{Collin_FieldTheoryOfGuidedWaves} or with high numerical precision \cite{Sihvola_2004,Helsing_2013}. The results are depicted in Fig.~\ref{fig1}.
\BF
\begin{center}
	\includegraphics[width=8.1cm]{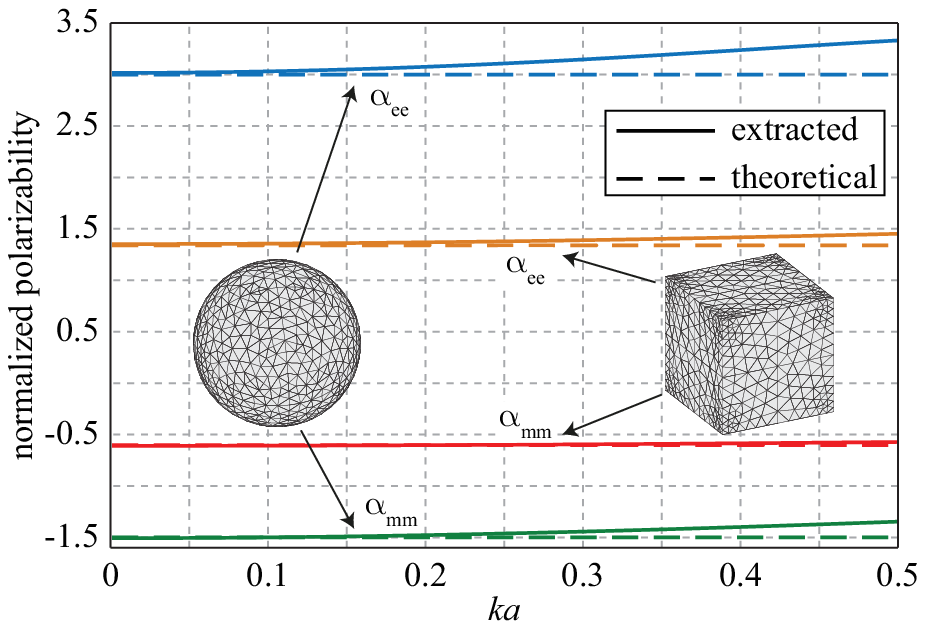}
	\caption{The real part of the normalized electric and magnetic polarizability of a PEC sphere and a PEC cube obtained by the method proposed in this paper. The results are compared to their known static values.  Note that the non-vanishing magnetic polarizability in the static limit is a consequence of using PEC. Even negligibly small losses will lead to $\alpha_{\mathrm{mm}}=0$ for $ka \to 0$, see \cite{Collin_FieldTheoryOfGuidedWaves,2014_Jelinek_PRB} and the references therein. }
	\label{fig1}
\end{center}
\EF
For small values of the normalized frequency $ka$, the correspondence of the presented method and analytical expressions for static polarizabilities is very good. Small discrepancies can be attributed to finite meshing. As the electrical size $ka$ increases, the discrepancy grows, but, in this case, one must realize that the formulations \cite{Collin_FieldTheoryOfGuidedWaves,Sihvola_2004,Helsing_2013} are strictly static and are not supposed to be precise for $ka>0$. Regarding the sensitivity of the results on the triangular mesh density, it is worth noting that, generally, the polarizability extraction is rather forgiving in this respect. For example, the difference between the extracted polarizability of a cube meshed by 100 triangles and the theoretical value at $ka=0.05$ is just a few percent for magnetic polarizability, as well as electric polarizability. The mesh refinement of the corners and edges also seems to play a minor role in the precision of the polarizability results.

The dynamic behaviour of polarizability presented in Fig.~\ref{fig1} could not be verified by the static analytical solution. It is, thus, of interest to compare it with the results of other methods. Such a comparison is made in Fig.~\ref{fig1a} where the polarizability extraction via the 2D periodic arrangement of scatterers \cite{ScherKuester2009,2012_Albooyeh_PRB,2013_Karamanos_IETMAP,2016_Liu_TAP} has been employed. The necessary interaction constant for the used square periodicity has been evaluated by analytical formulas shown in \cite{Tretyakov2003} (Sec.~4.5.2). Although the array method predicts the rise of polarizability with growing electrical size, the shape of this dependence is in considerable disagreement with the method of this paper. Furthermore, the array method shows an important dependence on the ratio of scatterer size to the array period which is a systematic error induced by a point dipole approximation of the interaction constant \cite{BelovSimovski2005,Tretyakov2003}. The method of Section~\ref{method} is free of this error as it naturally operates in free space.
\BF
\begin{center}
	\includegraphics[width=8.1cm]{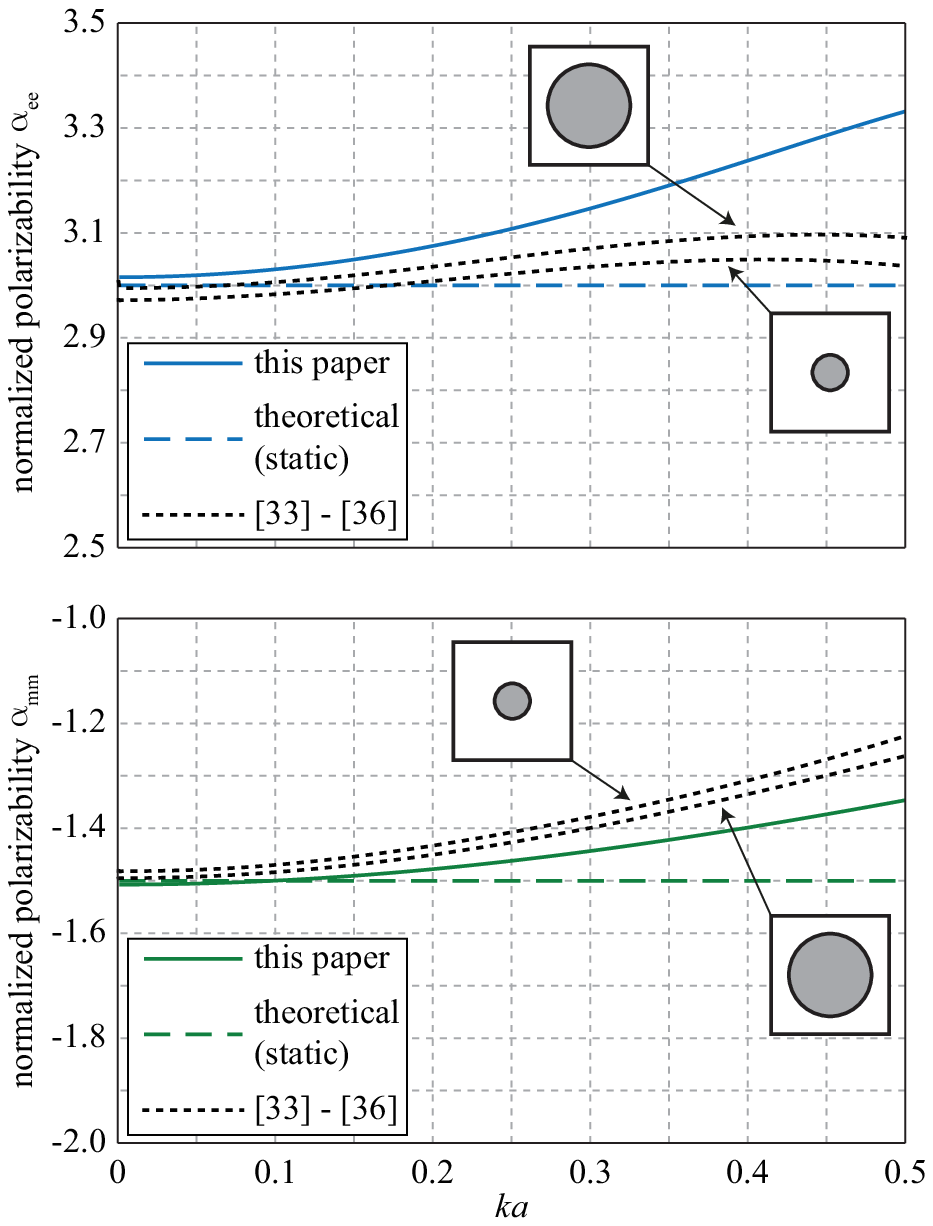}
	\caption{ The real part of the normalized electric and magnetic polarizability of a PEC sphere. The results of the method proposed in this paper are compared to their known static values as well as to the 2D array scattering technique \cite{ScherKuester2009,2012_Albooyeh_PRB,2013_Karamanos_IETMAP,2016_Liu_TAP}. The dependence of the latter on the ratio between the size of the sphere and the lattice period is shown. }
	\label{fig1a}
\end{center}
\EF

The full-wave formulation of the polarizability extraction method allows for a radiation correction, it, however, presents also its drawback when $ka \ll 1$ is desired.  The first issue comes from the frequency dependence of impedance matrix $\Tensor{\mathbf{Z}}$ (see Section~\ref{app}) which becomes ill-defined at $ka \to 0$. A second issue is caused by the Bessel-type excitation (see Section~\ref{excit}). In that case the sole excitation by magnetic field at $ka \to 0$ produces no excitation at all in the EFIE formulation (\ref{2Beq07}). The consequences of the two mentioned issues are illustrated in Fig.~\ref{fig2}.
\BF
\begin{center}
	\includegraphics[width=8.1cm]{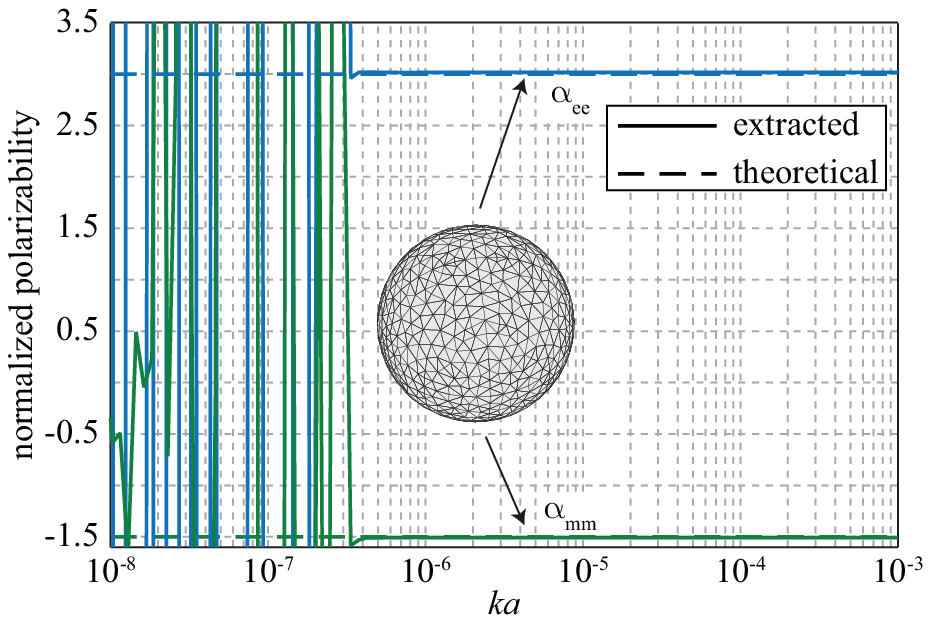}
	\caption{The real part of the normalized electric and magnetic polarizability of a PEC sphere. The results are compared to its known static values.}
	\label{fig2}
\end{center}
\EF
The ill-conditioning of impedance matrix $\Tensor{\mathbf{Z}}$ starts to be an issue below $ka=10^{-5}$ and the extraction method breaks down completely below $ka=10^{-6}$. Curiously, the Bessel-type excitation  presents no real problem (in the used double precision) in the $ka$ ranges allowed by the impedance matrix, which is seen from the fact that the extraction of the electric polarizability (not affected by the Bessel-type excitation) breaks at the same $ka$ as the extraction of the magnetic polarizability.

Encouraged by the good performance of the method on canonical objects, we can test it in more complicated scenarios where radiation effects become important. In that respect the PEC resonant scatterers of non-negligible electrical sizes are interesting testing grounds. As one example we have chosen the broadside-coupled split ring resonator (BCSRR) \cite{MarquesMedinaEl-Idrissi2002} extensively used in the design of magnetic metamaterials \cite{MarquesJelinekFreireEtAl2011}. The outline of the scatterer is depicted in Fig.~\ref{fig3}. Its normalized magnetic polarizability is well known \cite{MarquesMedinaEl-Idrissi2002,MarquesMesaMartelEtAl2003} and is given by
\BE
\label{3eq04}
\alpha _{{\mathrm{mm}}}^{zz} =\frac{{{\mu_0 \pi ^2}}}{V L}{\left( {{r_{{\mathrm{ext}}}} - \frac{w}{2}} \right)^4}{\left( {\frac{{\omega _0^2}}{{{\omega ^2}}} - 1 + {\mathrm{j}}\frac{{ {R_{{\mathrm{rad}}}}}}{{\omega L}}} \right)^{ - 1}},
\EE
where $r_{\mathrm{ext}}$ is the external radius of the ring, $w$ is the width of the strip, $L$ is the self-inductance of the resonator~\cite{MarquesMesaMartelEtAl2003}, $\omega_0$ is its resonance frequency and
\BE
\label{3eq05}
{R_{{\mathrm{rad}}}} = \frac{{{Z_0}\pi {k^4}}}{6}{\left( {{r_{{\mathrm{ext}}}} - \frac{w}{2}} \right)^4},
\EE
represents losses via radiation \cite{Tretyakov2003,Balanis_Wiley_2005}. The full-wave polarizability extracted via the method of this paper is compared to (\ref{3eq04}) in Fig.~\ref{fig3} for BCSRR of proportions $r_{{\mathrm{ext}}}/w=6$, $w=t$, where $t$ is the axial height of the resonator. An excellent agreement can be observed validating both the analytical model of \cite{MarquesMedinaEl-Idrissi2002} and the proposed method.
\BF
\begin{center}
	\includegraphics[width=8.1cm]{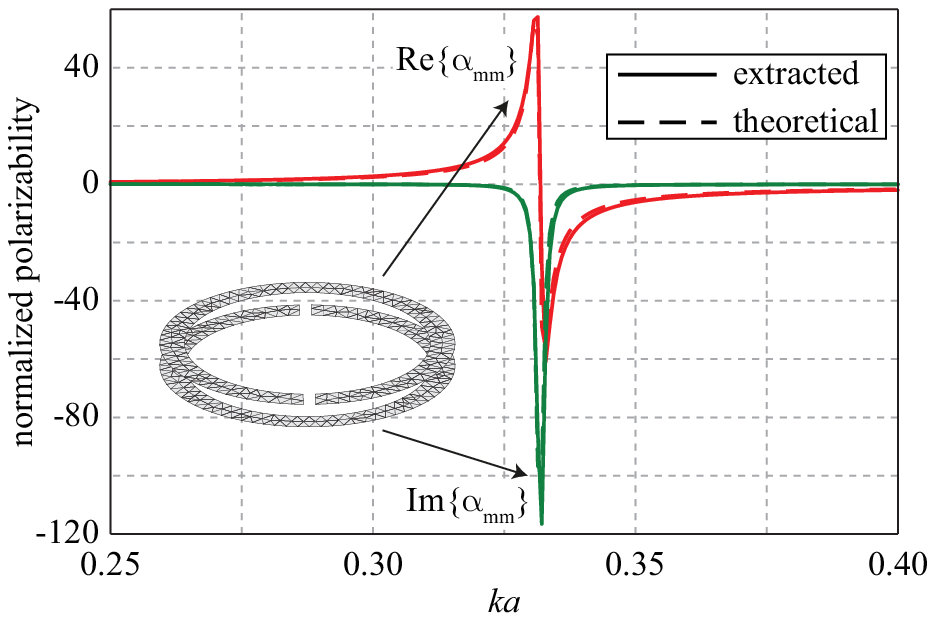}
	\caption{The real and imaginary part of the normalized magnetic polarizability of the BCSRR made of PEC. The results are compared to the analytical model proposed in \cite{MarquesMedinaEl-Idrissi2002}.}
	\label{fig3}
\end{center}
\EF

The BCSRR can also be used to test the inclusion of ohmic losses. To that point the analytical model (\ref{3eq04}) can easily be modified with ${R_{{\mathrm{rad}}}} \to {R_{{\mathrm{rad}}}} + {R_{{\mathrm{loss}}}}$, where
\BE
\label{3eq06}
{R_{{\mathrm{loss}}}} = \frac{{2\pi }}{{w\sigma \delta }}\left( {{r_{{\mathrm{ext}}}} - \frac{w}{2}} \right)
\EE
represents the conduction losses \cite{MarquesMesaMartelEtAl2003}, with $\sigma$ representing metal conductivity and \mbox{$\delta  = \sqrt {2/\left( {\omega \mu \sigma } \right)}$} representing penetration depth. The numerical comparison of the analytical model and the numerically extracted magnetic polarizability of lossy BCSRR is presented in Fig.~\ref{fig4}.
\BF
\begin{center}
	\includegraphics[width=8.1cm]{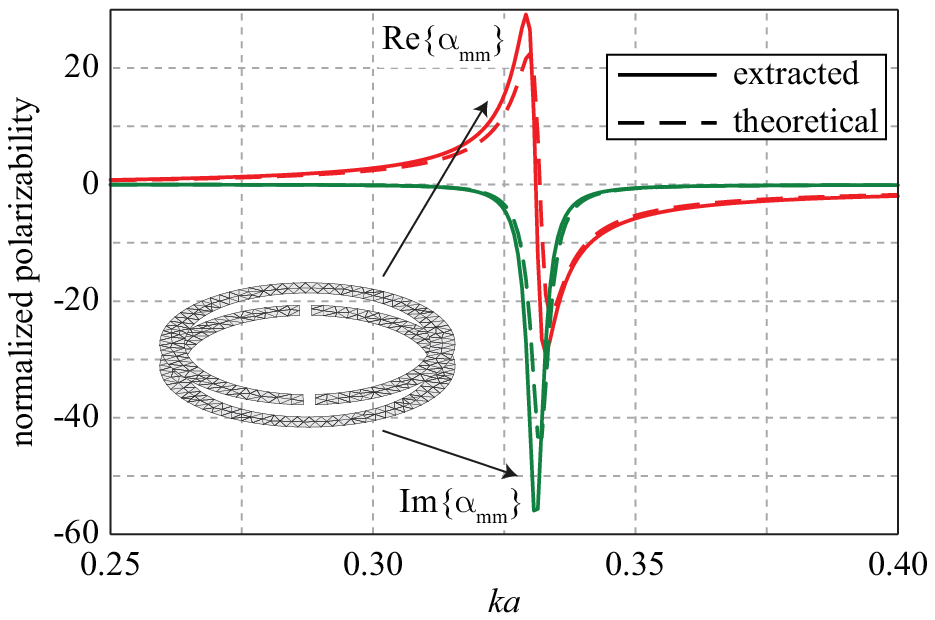}
	\caption{The real and imaginary part of the normalized magnetic polarizability of the lossy BCSRR. The results are compared to the analytical model proposed in \cite{MarquesMedinaEl-Idrissi2002}.}
	\label{fig4}
\end{center}
\EF
To keep the problem scalable, the ratio \mbox{$\sigma /  \left(\omega \epsilon_0\right)$} has been fixed at $10^7$, rather than the value of conductivity $\sigma$. The current layer thickness is assumed to be much bigger than the penetration depth, see Section~\ref{app}. Once more, excellent agreement is observed. Figures~\ref{fig3},~\ref{fig4} only show that the analytical model underestimates conduction losses (\ref{3eq06}). This is most probably caused by the non-negligible electrical size of the scatterer.

Based on the previous successful validations, we will now switch the paradigm, making the presented extraction technique a reference. In the last example of this paper the method will be used to investigate a precision of an analytical model of a complicated chiral scatterer \cite{MarquesJelinekMesa2007,JelinekMarquesMesaEtAl2008} called the chiral split ring resonator (ChSRR), whose geometry is depicted in Fig.~\ref{fig5}. An analytical model for its polarizabilities has been developed in \cite{MarquesJelinekMesa2007,JelinekMarquesMesaEtAl2008}. The works \cite{MarquesJelinekMesa2007,JelinekMarquesMesaEtAl2008} also propose proportions of the ChSRR with balanced normalized polarizabilities, i.e., with $\alpha_{\mathrm{mm}}=\alpha_{\mathrm{ee}}=\mathrm{j}\alpha_{\mathrm{em}}$, where all mentioned polarizabilities have axis--axis orientations, see \cite{MarquesJelinekMesa2007,JelinekMarquesMesaEtAl2008} for details.
\BF
\begin{center}
	\includegraphics[width=8.1cm]{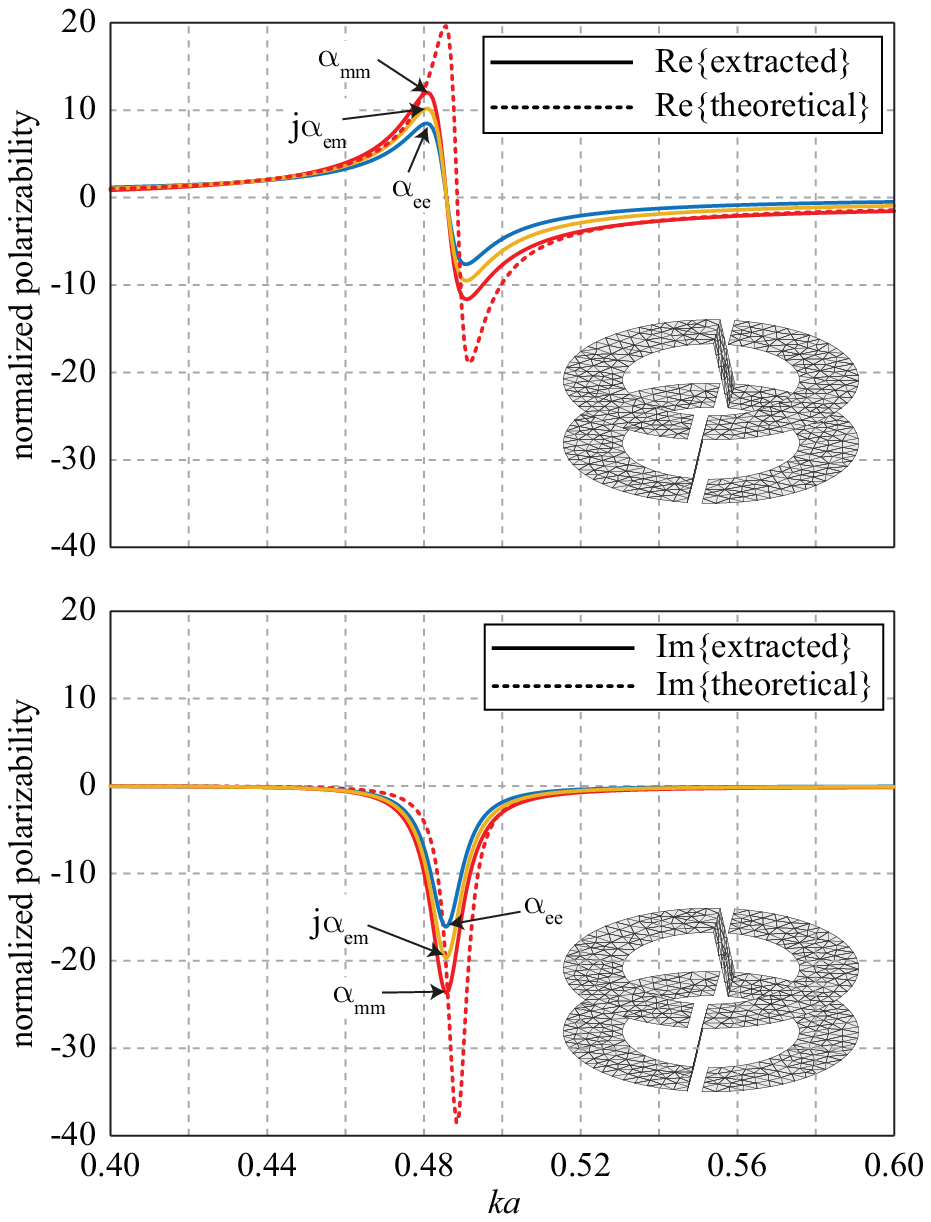}
	\caption{The real and imaginary part of the normalized polarizabilities of the ChSRR made of PEC. The results are compared to the analytical model proposed in \cite{MarquesJelinekMesa2007,JelinekMarquesMesaEtAl2008}. The particle is theoretically balanced,  i.e., $\alpha_{\mathrm{mm}}=\alpha_{\mathrm{ee}}=\mathrm{j}\alpha_{\mathrm{em}}=-\mathrm{j}\alpha_{\mathrm{me}}$.  Only one theoretical curve is thus depicted. The presented extraction method also automatically yields  $\alpha_{\mathrm{em}}=-\alpha_{\mathrm{me}}$. All depicted polarizabilities have axis--axis orientations, see \cite{MarquesJelinekMesa2007,JelinekMarquesMesaEtAl2008}.}
	\label{fig5}
\end{center}
\EF
The theoretical condition for a polarizability balance reads $2/\pi = k_0 r^2/  t$ \cite{MarquesJelinekMesa2007,JelinekMarquesMesaEtAl2008}, where $t$ is the thickness of the particle, $r$ is the mean radius of the ring and $k_0$ is the wavenumber at resonance given also by the width $w$ of the strip forming the resonator. Theoretically, this condition is satisfied by a particle of proportions $r/w \approx 2.0$, $r/t \approx 1.7$ which resonates at $k_0r \approx 0.38$. The precision of the analytical model is tested in Fig.~\ref{fig5}. Clearly, the relatively high electrical size of the particle leads to a serious underestimation of radiation losses by the analytical model. If the radiation loss is, however, fitted to the numerically extracted results (it is not relevant for periodic arrangements of the scatterers used for chiral metamaterial design \cite{JelinekMarquesMesaEtAl2008}), the correspondence becomes acceptable. Figure~\ref{fig5} also shows that polarizabilities are close to being balanced, although further optimization of the particle's dimensions would be necessary for a complete balance.

\section{Discussion}
\label{disc}
This section briefly recalls several important properties of the proposed polarizability extraction scheme, which could possibly be missed during the general exposition of the method:

\begin{itemize}

\item{The knowledge of polarizability tensors fully characterize the scattering properties of an electrically small scatterer. Namely, the differential scattering cross-section is defined as \cite{Bohren_scattering_by_Small_Particles,Jackson_ClassicalElectrodynamics}
\BE
\label{4eq01}
\sigma \left( {\theta ,\varphi } \right) = {r^2}\frac{{{{\left| {{\boldsymbol{E}}_{\mathrm{s}}^{{\mathrm{far}}}} \right|}^2}}}{{{{\left| {{{\boldsymbol{E}}_{\mathrm{i}}}} \right|}^2}}},
\EE
where ${{{{ {{\boldsymbol{E}}_{\mathrm{s}}^{{\mathrm{far}}}} }}}}$ is the scattered electric far-field \cite{Jackson_ClassicalElectrodynamics} and ${{{{ {{{\boldsymbol{E}}_{\mathrm{i}}}} }}}}$ is the electric field of the incident planewave. Assuming a small electrical size of the scatterer, the differential scattering cross-section can be written as \cite{Jackson_ClassicalElectrodynamics}
\BE
\label{4eq02}
\sigma \left( {\theta ,\varphi } \right) = {\left( {\frac{{{Z_0}{k^2}}}{{4\pi \left| {{{\boldsymbol{E}}_{\mathrm{i}}}} \right|}}} \right)^2}{\big| {{{\boldsymbol{r}}_0} \times \left( {c_0{\boldsymbol{p}} - {{\boldsymbol{r}}_0} \times {\boldsymbol{m}}} \right)} \big|^2},
\EE
where ${{\boldsymbol{r}}_0}$ is the unit vector in radial direction. The electric and magnetic dipole moments $\boldsymbol{p}$ and $\boldsymbol{m}$, corresponding to the exciting planewave with  ${{{{ {{{\boldsymbol{E}}_{\mathrm{i}}}} }}}}$, are, by means of (\ref{2Aeq01}), given by polarizability tensors.

}

\item{It can easily be checked that the reciprocity constraints \cite{LandauLifshitzPitaevskii1984}, namely, 
\BE
\label{4eq03}
\left[ {\begin{array}{*{20}{c}}
		{\boldsymbol{\Tensor{\alpha}}_{{\mathrm{ee}}}^{}}&{\boldsymbol{\Tensor{\alpha}}_{{\mathrm{em}}}^{}}\\
		{\boldsymbol{\Tensor{\alpha}}_{{\mathrm{me}}}^{}}&{\boldsymbol{\Tensor{\alpha}}_{{\mathrm{mm}}}^{}}
	\end{array}} \right]^{\mathrm{T}} = \left[ {\begin{array}{*{20}{c}}
	{\boldsymbol{\Tensor{\alpha}}_{{\mathrm{ee}}}^{}}&{-\boldsymbol{\Tensor{\alpha}}_{{\mathrm{em}}}^{}}\\
	{-\boldsymbol{\Tensor{\alpha}}_{{\mathrm{me}}}^{}}&{\boldsymbol{\Tensor{\alpha}}_{{\mathrm{mm}}}^{}}
\end{array}} \right]
\EE
are closely followed by the aforementioned implementation. This results from the implicit reciprocity within the used EFIE formulation which leads to symmetric matrix $\Tensor{\mathbf{Z}}$.

}

\item{Employing a volumetric version of the EFIE \cite{Wang_GeneralizedMoM}, the presented method can also be used in a straightforward way on scatterers containing dielectrics. The formulation presented in Section~\ref{method} would remain unchanged. The reformulation of the EFIE only changes matrices $\Tensor{\mathbf{Z}}$, $\Tensor{\mathbf{\Sigma}}$ and current densities $\boldsymbol{J}$ would need to be changed to polarization current densities $\mathrm{j} \omega \boldsymbol{P}$ in the dielectric regions, with $\boldsymbol{P}$ representing electric polarization.  There is, however, no simple way of modifying the method to account for non-reciprocal scatterers.}
\end{itemize}

\section{Conclusion}
\label{concl}
A full-wave method extracting all four polarizability tensors has been presented and tested on electrically small objects with known values of polarizability. Excellent agreement between numerical and analytical results has been observed both in the quasi-static and dynamic ranges. A noticeable merit of the presented scheme is the implicit inclusion of radiation and ohmic losses. 

The method is followed by a freely available implementation in Matlab working environment where the only user input is a triangular mesh of the scatterer's surface. The implementation has been enabled by the matrix formulation of the problem. The fast and effective evaluation of polarizability tensors allows for various optimization tasks concerning electrically small scatterers whose purpose can be found in the design of artificial media, radio identification tags and beam-forming arrays. Such an optimization can easily be performed via the modification of matrix $\Tensor{\mathbf{Z}}$ or, more simply, via the modification of matrix $\Tensor{\mathbf{\Sigma}}$ by locally varying the surface impedance.

%%%%%%%%%%%%%%%%%%%%%%%%%%%%%%%%%%%%%%%%%%%%%%%%%%%%%%%%%%%%%%%%%%%%%%%%%%%%%%%%%%%%%%%%%%%%%%%%%%%%%%%%%%%%%%%%%%%%%%%%%%%%%%%%%%%%%%%%%%%%%%%%%%%%%%%%%%%%%%%%%%%%%%%%%%%%%%

\section*{Acknowledgement}
This work was supported by the Czech Science Foundation under project No. 15-10280Y and project \mbox{No.~13-09086S}.

\section{Appendix}
\label{app}
The continuous form of the electric field integral equation (EFIE) \cite{Harrington_FieldComputationByMoM} used throughout this paper reads
\BE
\label{appA01}
{Z_{\mathrm{s}}}{\boldsymbol{K}}\left( {\boldsymbol{r}} \right) = -\boldsymbol{n}\left( {\boldsymbol{r}} \right) \times\boldsymbol{n}\left( {\boldsymbol{r}} \right) \times \Big( {\boldsymbol{E}}\left( {\boldsymbol{r}} \right) + {k^2}L\left\{ {{\boldsymbol{K}}\left( {\boldsymbol{r}} \right)} \right\} + \nabla L\left\{ {\nabla  \cdot {\boldsymbol{K}}\left( {\boldsymbol{r}} \right)} \right\} \Big)
\EE
with ${\boldsymbol{K}}\left( {\boldsymbol{r}} \right)$ being the surface current density induced on the scatterer, $ \boldsymbol{n}\left( {\boldsymbol{r}} \right)$ being the unit normal to the surface, ${\boldsymbol{E}}\left( {\boldsymbol{r}} \right)$ being the incident electric field and with operator $L$ defined as
\BE
\label{AppA02}
L\left\{{\boldsymbol{F}}\left( {\boldsymbol{r}} \right)\right\} = \frac{{ - \mathrm{j}}}{{4\pi \omega \varepsilon }} \int\limits_{S'} {{\boldsymbol{F}}\left( {{\boldsymbol{r'}}} \right)\frac{{{{\mathrm{e}}^{ - {\mathrm{j}}k\left| {{\boldsymbol{r}} - {\boldsymbol{r'}}} \right|}}}}{{\left| {{\boldsymbol{r}} - {\boldsymbol{r'}}} \right|}}{\mathrm{d}}S'}.
\EE
The quantity \mbox{${Z_{\mathrm{s}}} = \left( {1 + {\mathrm{j}}} \right)/\left( {\sigma \delta } \right)$} represents the surface impedance of the conducting half-space \cite{Jackson_ClassicalElectrodynamics} with \mbox{$\delta  = \sqrt {2/\left( {\omega \mu \sigma } \right)}$} representing penetration depth. The LHS of (\ref{appA01}) thus approximates the reaction of a lossy conductor in cases where the penetration depth is negligible with respect to thickness and with respect to the curvature radius of any part of the scatterer.
Utilizing expansion (\ref{2Beq03}), the integral equation (\ref{appA01}) is recast into its matrix form (\ref{2Beq04}),
where
\BE
\label{AppA05}
\Tensor{\mathbf{Z}} = \Big[ {\left( {{k^2}\left\langle {{{\boldsymbol{f}}_m},L\left\{{{\boldsymbol{f}}_n}\right\}  } \right\rangle  - \left\langle {\nabla  \cdot {{\boldsymbol{f}}_m},L\left\{\nabla  \cdot {{\boldsymbol{f}}_n}\right\} } \right\rangle } \right)} \Big]
\EE
is the so-called impedance matrix \cite{Harrington_FieldComputationByMoM,Harrington_MatrixMethodsForFieldProblems} and where
\BE
\label{AppA06}
{\Tensor{\mathbf{\Sigma }}} = {Z_{\mathrm{s}}}\Big[ {\left\langle {{{\boldsymbol{f}}_m},{{\boldsymbol{f}}_n}} \right\rangle } \Big]
\EE
is the matrix representing the reaction of a lossy conductor.

Effective ways of evaluating the matrix terms in (\ref{AppA05}) have been proposed by many authors \cite{Gibson_MoMinElectromagnetics}. In this paper, and in the code developed along \cite{polarizability_extraction_at_file_exchange}, we utilize the scheme of Makarov \cite{Makarov_AntennaAndEMModelingWithMatlab} which applies to the RWG basis \cite{RaoWiltonGlisson_ElectromagneticScatteringBySurfacesOfArbitraryShape}. The works \cite{RaoWiltonGlisson_ElectromagneticScatteringBySurfacesOfArbitraryShape,Makarov_AntennaAndEMModelingWithMatlab} also contain an effective evaluation scheme for the RHS of (\ref{2Beq04}).

\subsection{Surface resistivity matrix}
\label{Sigmamat}
The surface resistivity matrix ${\Tensor{\mathbf{\Sigma }}}$ is presented in this subsection in its explicit form within the RWG basis as, to the best of the authors' knowledge, it cannot be found elsewhere. The evaluation of the scalar products (\ref{AppA06}) within the RWG basis is performed in barycentric coordinates and leads, after some relatively straightforward algebra, to 
\BE
\label{AppD03}
\begin{aligned}
\left\langle {{{\boldsymbol{f}}_m},{{\boldsymbol{f}}_m}} \right\rangle  &= \frac{{l_m^2}}{{24A_m^ + }}\Bigg[ {\boldsymbol{r}}_m^{\left( {{\mathrm{c}+}} \right)} \cdot \left( {9{\boldsymbol{r}}_m^{\left( {{\mathrm{c}+}} \right)} - 15{\boldsymbol{v}}_m^{\left( 1 \right)}} \right) \\
&+ 7{{\left| {{\boldsymbol{v}}_m^{\left( 1 \right)}} \right|}^2} - {\boldsymbol{v}}_m^{\left( 2 \right)} \cdot {\boldsymbol{v}}_m^{\left( 3 \right)} \Bigg] + \\
&+ \frac{{l_m^2}}{{24A_m^ - }}\Bigg[ {\boldsymbol{r}}_m^{\left( {{\mathrm{c}}-} \right)} \cdot \left( {9{\boldsymbol{r}}_m^{\left( {{\mathrm{c}}-} \right)} - 15{\boldsymbol{v}}_m^{\left( 4 \right)}} \right) \\
&+ 7{{\left| {{\boldsymbol{v}}_m^{\left( 4 \right)}} \right|}^2} - {\boldsymbol{v}}_m^{\left( 2 \right)} \cdot {\boldsymbol{v}}_m^{\left( 3 \right)} \Bigg]
\end{aligned}
\EE
for diagonal terms and
\BE
\label{AppD04}
\begin{aligned}
\left\langle {{{\boldsymbol{f}}_m},{{\boldsymbol{f}}_n}} \right\rangle  &= \frac{{{\chi _{mn}}{l_m}{l_n}}}{{24{A_m}}}\Bigg[ 9{\boldsymbol{r}}_m^{\left( {\mathrm{c}} \right)} \cdot \left( {{\boldsymbol{r}}_m^{\left( {\mathrm{c}} \right)} - {\boldsymbol{v}}_m^{\left( {\mathrm{f}} \right)} - {\boldsymbol{v}}_n^{\left( {\mathrm{f}} \right)}} \right) \\
&+ {{\left| {{\boldsymbol{v}}_m^{\left( {\mathrm{f}} \right)} + {\boldsymbol{v}}_n^{\left( {\mathrm{f}} \right)}} \right|}^2} + 5{\boldsymbol{v}}_m^{\left( {\mathrm{f}} \right)} \cdot {\boldsymbol{v}}_n^{\left( {\mathrm{f}} \right)} \Bigg]
\end{aligned}
\EE
for off-diagonal terms, with $l_m$ as the edge length of the \mbox{$m$-th} RWG function, ${A_m^{\pm} }$ as the area of its \mbox{positive / negative} triangle and ${\boldsymbol{r}}_m^{\left({\mathrm{c}} \pm \right)}$ as the \mbox{positive / negative} triangle centre
\cite{RaoWiltonGlisson_ElectromagneticScatteringBySurfacesOfArbitraryShape}. The vertices $\boldsymbol{v}$ are defined according to Fig.~\ref{fig14}a. The \mbox{superindex $^{\left( {\mathrm{f}} \right)}$}, used in (\ref{AppD04}), denotes free vertices (the vertices $\boldsymbol{v}^{\left( 1 \right)}$ and $\boldsymbol{v}^{\left( 4 \right)}$) belonging to the triangle common to the \mbox{$m$-th} and the \mbox{$n$-th} RWG function. The coefficient $\chi_{mn}$ is equal to unity for cases depicted in Fig.~\ref{fig14}b,c, to minus unity for cases depicted in Fig.~\ref{fig14}d,e and to zero for RWG functions with no common triangle.
\BF
\begin{center}
	\includegraphics[width=8.1cm]{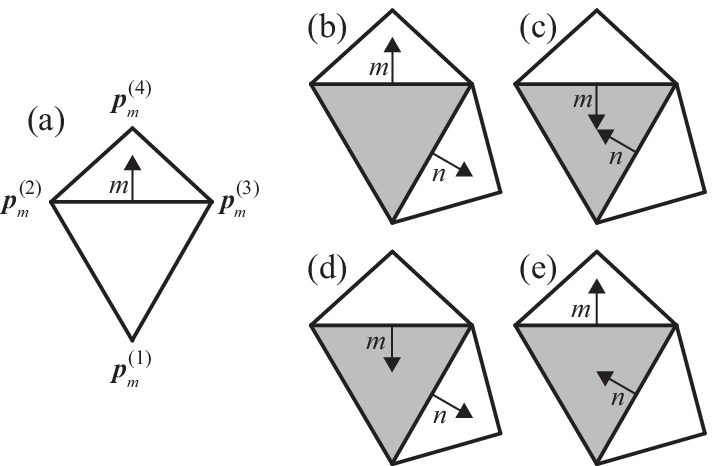}
	\caption{Sketch of the $m$-th RWG function (a) and of an overlap (b, c, d, e) between the $m$-th and the $n$-th RWG function. The orientation of the RWG function is denoted by an arrow. The vertices are denoted by corresponding radius vectors $\boldsymbol{v}_m$. The grey colour represents the overlap region.}
	\label{fig14}
\end{center}
\EF

\subsection{Electric and Magnetic Dipole Matrices}
\label{PMmatrices}
Though elementary, the explicit forms of matrices $\Tensor{\mathbf{P}}$ and $\Tensor{\mathbf{M}}$ in the RWG basis are presented in this subsection, allowing readers to implement the polarizability extraction method presented in this paper directly. The derivation starts with the substitution of (\ref{2Beq03}) into relations (\ref{2Aeq02a}) and (\ref{2Aeq02b}) defining the electric and magnetic dipole moments. By then performing a direct integration in barycentric coordinates, this leads to
\BE
\label{AppE01}
{\Tensor{\mathbf{P}}} = \frac{1}{{{\mathrm{j}}\omega }}\left[ {\begin{array}{*{20}{c}}
	{{l_1}\left( {{\boldsymbol{r}}_1^{\left( {{\mathrm{c}} - } \right)} - {\boldsymbol{r}}_1^{\left( {{\mathrm{c}} + } \right)}} \right)}& \cdots &{{l_N}\left( {{\boldsymbol{r}}_N^{\left( {{\mathrm{c}} - } \right)} - {\boldsymbol{r}}_N^{\left( {{\mathrm{c}} + } \right)}} \right)}
	\end{array}} \right]
\EE
and
\BE
\label{AppE02}
{\Tensor{\mathbf{M}}} = \frac{1}{4}\left[ {\begin{array}{*{20}{c}}
	{{l_1}\left( {{\boldsymbol{v}}_1^{\left( 1 \right)} \times {\boldsymbol{r}}_1^{\left( {c + } \right)} - {\boldsymbol{v}}_1^{\left( 4 \right)} \times {\boldsymbol{r}}_1^{\left( {c - } \right)}} \right)}& \cdots &{{l_N}\left( {{\boldsymbol{v}}_N^{\left( 1 \right)} \times {\boldsymbol{r}}_N^{\left( {c + } \right)} - {\boldsymbol{v}}_N^{\left( 4 \right)} \times {\boldsymbol{r}}_N^{\left( {c - } \right)}} \right)}
	\end{array}} \right],
\EE
where the vertices $\boldsymbol{v}$ are defined according to Fig.~\ref{fig14}a. As a word of caution we mention that matrix ${\Tensor{\mathbf{M}}}$ is generally coordinate dependent, since the divergence of the surface current density is not vanishing \cite{LandauLifshitzPitaevskii1984,Merlin2009}. Displacement of the coordinate center along a constant vector $\mathbf{d}$, results in a change
\BE
\label{AppE03}
{{\Tensor{\mathbf{M}}}\mathbf{I}} \to {{\Tensor{\mathbf{M}}}\mathbf{I}} + \frac{{{\mathrm{j}}\omega }}{2}{\boldsymbol{d}} \times \left( {{{\Tensor{\mathbf{P}}}\mathbf{I}}} \right).
\EE
Keeping the vector $\boldsymbol{d}$ within the scatterer makes the ambiguity of the order of $ka$ which was assumed to be small.

\ifCLASSOPTIONcaptionsoff
  \newpage
\fi

%\bibliographystyle{IEEEtran}
%\bibliography{e:/Google_Drive/Lukas/!!!TEMPLATES/references_LIST_UpToDate}
% Generated by IEEEtran.bst, version: 1.13 (2008/09/30)

\end{document}